# Dynamically Switchable Polarization Lasing between q-BIC and Bragg Resonance Modes

Hongyu Yuan, Jiaoyao Liu, Xiaolin Wang, Qianwen Jia, Jinwei Shi, Dahe Liu and Zhaona Wang[*]

Key Laboratory of Multiscale Spin Physics (Ministry of Education), Applied Optics Beijing Area Major Laboratory, School of Physics and Astronomy, Beijing Normal University, Beijing 100875, China

[*]*Contact author: zhnwang@bnu.edu.cn*

**Abstract**

Quasi-bound states in the continuum (q-BICs) enable low-threshold lasing through high-$Q$ cavity modes, yet their polarization tunability remains constrained by nanostructure-imposed cavity symmetries. By engineering a microcavity with an optimized duty cycle (0.34), we demonstrate a polarization-switchable distributed feedback (DFB) laser with controlled emission transitions between dual off-Γ q-BIC lasing and single Γ-point Bragg resonance (BR) lasing through switching pump polarization. The switching mechanism shows unprecedented robustness in varying waveguide thickness and photonic crystal period of DFB structures. Our findings extend the capabilities of DFB lasers beyond their conventional limits, opening opportunities for nanophotonics, classical and quantum optics applications.

## Introduction

Microcavity lasers have significant application potential in the fields of lighting and display, communications, and optoelectronic integration due to their advantages of compact size and ease of integration. Generally, the resonance modes within the cavities of Fabry–Perot (FP)[1], whispering-gallery[2], and distributed feedback (DFB)[3] are utilized to achieve lasing under a low excitation threshold. Among them, the DFB microcavities have been widely fabricated through grating[4,5] or photonic crystal (PC)[6] with periodic refractive index modulation to create multiple scattering and feedbacks. The DFB laser is typically characterized with a narrow linewidth in spectra, and the far-field radiation pattern of "bright spot" at the high-symmetry points[7-9]. The emission modes satisfy Bragg resonance (BR) condition as a BR lasing. However, there are still some reports on special radiation patterns of "double bright lines" from one-dimensional (1D) PC and "doughnut-shape" from two-dimensional (2D) PC at back focus plane[10,11]. The absence of radiation at high-symmetry points in DFB lasers was looked as band-edge lasing[12,13]. These two types of laser emissions occur in different systems respectively, indicating that the key parameters determined the radiation mode of DFB lasers are still open questions to be uncovered through high-resolution angular spectrum characterization techniques.

As an emerging resonant mode in PC, the bound states in the continuum (BICs) have been utilized to design and realize the quasi-BIC (q-BIC) lasers with extremely low threshold and unique momentum-space radiation characteristics through breaking symmetry in real or parameter space[14-16]. From the perspective of feedback mechanisms, the q-BIC lasers based on PC cavity inherently belong to the DFB lasers due to the periodic scattering feedback[17]. The BIC stems from the strong coupling between the electromagnetic modes, leading to a symmetric mode as dark mode (i.e., BIC) without radiative loss and an anti-symmetric mode as bright mode with radiative loss[18,19]. Due to the non-radiative nature of BICs, q-BIC lasers exhibit unique emission patterns with center dark at the BIC resonance conditions. These patterns include "double bright lines"[20,21], "doughnut-shape"[17,22-26], and several radiation lobes shaped by lattice symmetry[27,28]. Currently, extensive efforts are devoted to realizing q-BIC laser due to its clear mechanism and low threshold[29]. In contrast, the lasing behavior of bright modes near the BR condition has rarely been reported. This is primarily constrained by the conventional viewpoint that bright modes are difficult to utilize for achieving laser emission due to their relatively high radiative losses[30]. Recently, center-bright emission patterns at high-symmetry points corresponding to bright modes are observed in a special symmetry-broken 2D PC microcavity[24] and a finite-sized plasmonic lattice[31], indicating the possibility of BR lasing. How to achieve BR lasing based on the bright mode by regulating parameters of PC cavities? Can dynamic switching lasing between q-BIC and BR modes be achieved? These questions remain critical yet unresolved in the field of microcavity lasers.

Here, we selectively excite the q-BIC and BR modes using transverse electric (TE)

or transverse magnetic (TM) polarized waves, respectively. This is achieved by tuning TE-wave Bloch modes into the strong-coupling regime and TM-wave Bloch modes into the weak-coupling regime through manipulating the duty cycle (DC) of the PC. A polymer DFB laser with a PC-waveguide-coupled structure is fabricated via a single-step interference lithography at an optimized DC and is characterized by a custom high-resolution angle-resolved imaging and spectroscopic system, enabling precise control of structural parameters and accurate discrimination of lasing modes in momentum space. Furthermore, we experimentally demonstrate dynamic transitions between q-BIC lasing and BR lasing by adjusting the pump polarization under varying waveguide thickness and PC period at a fixed DC.

**Results**

**The effect of DC on PC-waveguide-coupled structure**

A polymer DFB laser was designed as the 1D PC-waveguide-coupled structure shown in Fig. 1(a). In the coupled structure, the electromagnetic mode of the 1st-order diffraction that satisfy the Bloch condition along $x$ direction are quantized into different branches of $m$-th Bloch mode named as $TE_m^1$ ($m$=0,1,2,……) for TE waves and $TM_m^1$ ($m$=0,1,2,……) for TM waves due to the resonance modification of the waveguide in the vertical direction (Section I, Supplemental Material)[32]. The positive and negative 1st-order diffraction electromagnetic modes couple at the points satisfying the generalized BR condition at the interface. The mode coupling leads to a dark mode $TE_{d,m,m'}$ and a bright mode $TE_{b,m,m'}$ ($m$ or $m'$=0,1,2,……) for TE waves, and $TM_{d,m,m'}$ and $TM_{b,m,m'}$ for TM waves near the BR condition for the coupled structures with different DC (Fig. S1, Supplemental Material). When the Bloch modes share the same order ($m=m'$), the resulting dark mode manifests as a symmetry-protected BIC (SP-BIC)[33]. Conversely, when inter-order Bloch mode coupling ($m\neq m'$) occurs, the system exhibits a Friedrich-Wintgen BIC (FW-BIC)[34].

To uncover the localization property of the dark mode in momentum space, the full width at half maximum (FWHM) of the quality ($Q$) factor curve of the resonance modes along the ΓX direction is defined as the angular width $\Delta\alpha$ and calculated through interpolation fitting. Here, $\sin\alpha$ represents the normalized parallel wave vector component and $\alpha$ is the incident angle. The variations of $\Delta\alpha$ for the TE and TM resonance modes demonstrate two obvious regions through varying DC at a fixed PC period of 380 nm (Fig. 1(b)). When the DC exceeds 0.41 (e.g. the DC is 0.53 in Fig. S1, Supplemental Material ), the angular widths $\Delta\alpha$ of the $TM_{d,1,1}$ and $TE_{d,1,1}$ modes are relatively large and comparable, indicating a similar localization behavior of the BIC modes for TE waves and TM waves. The $Q$ values for the corresponding q-BIC modes are also huge for both TE and TM waves, meaning a flexible way to achieve different q-BIC lasing modes[20,21]. In contrast, when the DC is smaller than 0.41 (e.g. the DC is 0.29 in Fig. S1, Supplemental Material), the angular width $\Delta\alpha$ of the $TM_{d,1,1}$ mode is greatly smaller than that of the $TE_{d,1,1}$ mode. The $TM_{d,1,1}$ mode is strongly localized at the Γ point in momentum space, indicating the corresponding q-BIC mode possessing large irradiation loss, which is not benefit for achieving q-BIC lasing. For the bright modes, the $Q$ value of the $TM_{b,1,1}$ is always larger than that of the $TE_{b,1,1}$ due to the BR-induced leaky mode[35]. Therefore, when the DC is below 0.41, the BR (q-BIC) mode associated with the $TM_{b,1,1}$ ($TE_{d,1,1}$) can be excited by TM (TE) polarized waves.

**The polarization-switchable mechanism in PC-waveguide-coupled structure**

In our experiments, the DFB microcavity was fabricated through two-beam interference method with a tunable DC controlled by the development period. The PC-waveguide-coupled structure with a DC of 0.34 was achieved as shown in Fig. S2 (Supplemental Material). An angle-resolved spectra and imaging measurement system with extremely high angle resolution capability as a minimal angle spacing of 0.002 rad is home-made to reveal the coupling role between different modes near BR condition (Fig. S3, Supplemental Material). The angle-resolved transmission spectrum in Fig. 1(c) demonstrates the dispersion property of the Bloch modes, which show excellent agreement with the simulated results through eigenmode analysis (white dashed lines). The resonant transmission modes are observed near the Γ point for TE (TM) waves generated by the coupling of the $TE_1^1$ and $TE_1^{-1}$ ($TM_1^1$ and $TM_1^{-1}$) modes marked with rectangle. Two distinct resonance peaks were observed with a splitting interval ($\Delta\lambda$) of about 2.286 nm exceeding the resonance linewidth ($\delta\lambda=2.041$ nm)) for TE waves, indicating a strong coupling effect between the two resonant modes inducing an anti-crossing ($g$=4.05 meV)[36,37] in the energy bands (Fig. S4, Supplemental Material ). With increasing incident angle, the resonance peak at 605.398 nm (607.684 nm) exhibits a blueshift (redshift), indicating the mode located at the high-energy (low-energy) band branch in wavelength space. The corresponding local dispersion curves along the ΓX direction (Fig. 1(d)) also illustrate an obvious band splitting induced by strong coupling between the $TE_1^1$ and $TE_1^{-1}$ modes. The dark (bright) mode is located at the Γ point within the high-energy (low-energy) band branches, further confirmed by the antisymmetric (symmetric) field profiles due to non-radiative (radiative) nature (Fig. 1(f)). So, we conclude that the resonance peak at 605.398 nm (607.684 nm) is the q-BIC (bright) mode. In contrast, only a single resonance peak was obviously detected for TM waves, demonstrating a smaller splitting interval (Δλ) than the resonance linewidth due to weak coupling dominance [38]. The corresponding BR mode emerges as the dominant behavior due to limited energy transfer between the resonant modes. The resonance peak at 605.993 nm displays a redshift with increasing incident angle, indicating a bright mode emission at the upper band branch for TM waves. The results unambiguously demonstrate polarization-controlled band engineering in the designed photonic systems through selective activation of q-BIC and BR modes at the high-symmetry Γ point.

The Q-factor of high-Q modes in the ΓX direction (Fig. 1(e)) reveals that the dark mode of $TE_{d,1,1}$ has relatively large coupling region in momentum space, which is benefit for q-BIC mode excitation for a finite-sized photonic structure[39]. In contrast, the bright mode of $TM_{b,1,1}$ demonstrate substantially high Q factors and a relatively strong localization in momentum space, supplying a narrow leaky channel originating from guided-mode resonance (GMR) effects[40]. The unique polarization-controlled band splitting characteristic of the PC-waveguide-coupled structure enable a switchable q-BIC lasing with symmetric radiation profiles flanking the Γ point and BR lasing featuring emission at the Γ point from this designed DFB microcavity. The similar behaviors can also be found at the off-Γ position with sin$α$ = 0.017 for TE waves (green pentagram) and sin$α$ = 0.019 for TM waves (brown pentagram, Fig.1(c)) due to inter-order Bloch mode coupling. A redshift (dλ/d$θ$=5.188 nm/deg) and a concurrent blueshift (dλ/d$θ$=−2.080 nm/deg) are revealed in TE-polarized transmission spectra, providing a direct evidence of strong coupling ($g$=4.35 meV). The local dispersion curves along the ΓX direction near the off-Γ high-symmetry position exhibit the similar polarization-dependent band splitting behavior (Fig. 1(g)). A relatively large band splitting is observed at sin$α$ = 0.017 for TE waves, exhibiting a dark (bright) mode characteristic through the electromagnetic field distributions (Fig. 1(i)). The Q factor slowly drop away from the high-symmetry point in Fig. 1(h), enabling a q-BIC mode emission for a finite-size structure. In contrast, the resonance coupling is observed at $\sin\alpha = 0.019$ as the theoretical prediction from BR resonance conditions for TM waves (Fig. S1, Supplemental Material). The corresponding dark mode $TM_{b,0,1}$ is extremely sensitive to symmetry-breaking perturbations as its $Q$ factor drops sharply away from the high-symmetry point[39]. And a monotonic redshift of the resonance wavelength ($d\lambda/d\theta$=6.368 nm/deg) is for TM waves, indicating the bright mode property. Consequently, $TE_{d,0,1}$ mode and $TM_{b,0,1}$ mode can be used to achieve q-BIC lasing and BR lasing for the finite-sized sample even at the off-$\Gamma$ point.

**The polarization-switchable behaviors in the PC-waveguide-coupled laser**

The emission properties of this PC-waveguide-coupled DFB microcavity with a DC of 0.34 were systematically characterized using angular-resolved photoluminescence (PL) spectroscopy under nanosecond pulsed laser excitation. Under the TE polarization excitation, a distinct grating diffraction pattern was observed at a low pump energy density (Fig. 2(a)). When pump energy density increases (Fig. 2(b)), characteristic two-lobe patterns condensed near the $\Gamma$ point and at $\sin\alpha = \pm 0.017$ (white dashed-line box), indicating the formation of two q-BIC lasing modes corresponding to the SP-BIC at the $\Gamma$ point and the FW-BIC from inter-order Bloch mode coupling ($m$=0 and $m'$=1) at the off-$\Gamma$ point. The lasing behavior was also evidenced by the lasing profile in real space (left inset), and localized lasing pattern in momentum space (right inset, NA=0.023). The q-BIC lasing at the $\Gamma$ point exhibits an emission wavelength of 605.176 nm and a FWHM of 0.11 nm, yielding a $Q$ factor of approximately 5502 (Fig. 2(e)). This $Q$ factor represents 2.75 times higher than the previously reported value in a similar q-BIC laser[20]. The threshold of this q-BIC lasing is 0.412 mJ/cm², as characterized in the nonlinear behavior of integrated intensity (Fig. 2(f)) and decrease in the linewidth (Fig. 2(f)). This finding demonstrates that q-BIC lasing is realized under TE polarization pump using a full polymer DFB microcavity with a low-refractive-index of 1.64, which is the lower than the previous reports on q-BIC lasing[14,41,42].

For TM polarization excitation, a bright spot at the $\Gamma$ point emerges under a high pump energy density exceeding the threshold of 0.475 mJ/cm² (Fig. 2(d)), contrasting with the emission pattern under below-threshold pumping (Fig. 2(c)). The emission with a central intensity peak in momentum space (left inset) confirms the origin of BR lasing. The obtained BR lasing has a wavelength of 604.849 nm (Fig. 2(e)) and a

FWHM of 0.23 nm. The $Q$ factor is about 2630, slightly lower than that of the q-BIC lasing. The experimental results demonstrate a viable approach to realizing BR lasing under the TM polarization pump in this DFB laser.

The dynamic switching of the q-BIC and BR lasing was visualized through real-time momentum-space radiation patterns (video, Supplemental Material) and corresponding PL spectra (Fig. 2(g)) under dynamically modulated pump polarization. Under TE wave excitation with polarization angle $\theta_p=0°$ relative to grating orientation, two discrete lasing modes were observed from two distinct q-BIC states, as mentioned in the aforementioned Fig. 2(b). The q-BIC lasing near the Γ point exhibited a linear radiation polarization parallel to the grating lines with a degree of 0.94 (Fig. 2(h)). As the pump polarization angle $\theta_p$ increases, we observe a polarization-controlled mode collapse wherein the dual-mode q-BIC lasing degenerates into single-mode operation with gradually diminishing radiation intensity due to the pump efficiency reduction of TE waves (upper panel in Fig. 2(i)). Under complete TM polarization pump ($\theta_p=90°$), the q-BIC lasing was fully quenched, while BR lasing prevailed as the sole dominant emission mechanism, exhibiting orthogonal linear polarization with a degree of 0.96 relative to the grating lines (Fig. 2(h)). Notably, at an intermediate pump polarization angle of $\theta_p=80°$, two orthogonal lasing modes coexist, characterized by diminishing q-BIC emission alongside emerging BR lasing, with polarization-resolved measurements revealing their distinct origins (lower panel in Fig. 2(i)). The dynamical switching phenomenon demonstrates complete polarization control over the lasing mode selection, enabling reliable switching between q-BIC and BR lasing through pump polarization tuning from TE to TM configurations.

**Robustness of polarization-switchable behaviors in the DFB lasers**

To further demonstrate the robustness and scalability of this mode-switchable laser scheme, the PC-waveguide-coupled DFB lasers with different waveguide thicknesses at a fixed DC of 0.34 were designed. The $Q$ factor variations of the resonance modes along the ΓX direction exhibits similar polarization-controlled behaviors in angular width (Fig. S5, Supplemental Material), theoretically demonstrating the feasibility of lasing modes switching. The band structures of the fabricated DFB lasers also demonstrate distinct behaviors of strong couple for TE waves and weak couple for TM waves (Fig. 3(a-c)). The corresponding emission lasing spectra (Fig. 3(d)) are tuned by the pump polarization of TE and TM waves, indicating the capability of pump polarization to switch lasing modes. This mode-switchable behavior manifests as q-BIC lasing modes under TE polarization excitation and BR lasing modes under TM polarization excitation, as further verified by the corresponding radiation patterns in the coupling region of momentum space (inset in Fig. 3(d)). These results demonstrate the strong adaptability of the scheme to switch lasing modes via pump polarization in DFB lasers with different waveguide thicknesses.

In addition, increasing waveguide thickness enables the transition from single-mode to multi-mode lasing as verified in Fig. 3(d). For the DFB laser with a small thickness of 0.90 μm (red spectrum), single-mode q-BIC lasing and single-mode BR lasing is obtained from the 0th-order Bloch modes coupling under TE and TM waves excitation, respectively. As the waveguide thickness increases to 1.65 μm (cyan spectrum), multiple lasing modes emerge. The q-BIC lasing emerges via dark modes ($TE_{d,1,1}$ near Γ-point, $TE_{d,0,1}$ at $\sin\alpha = 0.017$, and $TE_{d,1,2}$ at $\sin\alpha = 0.028$), while BR lasing occurs through bright modes ($TM_{b,1,1}$ at Γ-point, $TM_{b,0,1}$ at $\sin\alpha = 0.019$, and $TM_{b,1,2}$ at $\sin\alpha = 0.030$). In the structures with thicknesses above 1.90 μm, higher-order Bloch modes dominate the coupling behavior, exhibiting the q-BIC lasing induced by dark modes $TE_{d,2,2}$ near the Γ point and $TE_{d,1,2}$ at $\sin\alpha = 0.030$, accompanied by bright-state radiation at both Γ point and $\sin\alpha = 0.032$. The observed radiation angle differences between q-BIC lasing and BR lasing show excellent agreement with dispersion relation (white dashed lines in Fig. 3(a-c)), uncovering the origin from inter-order Bloch mode coupling ($m \neq m'$). Notably, varying waveguide thickness not only maintains the polarization-switchable mechanism in DFB laser but also enables dual-wavelength and even multi-wavelength lasing.

The polarization-switchable behavior is equally applicable to the PC-waveguide-coupled DFB lasers with different PC periods, as predicted by their universal $Q$-factor angular width trend for TE and TM waves (Fig. S6, Supplemental Material). The DFB lasers with different PC periods demonstrates single-mode q-BIC lasing (solid lines) and single-mode BR lasing (dashed lines) under the TE and TM polarizations pump, respectively. The emission wavelength is modified from 578 nm to 618 nm through controlling the period $\Lambda$ (Fig. 4(a)), indicating the excellent wavelength-tunability as single-mode laser. The measured wavelength of the BR lasing (blue dots) exhibits a linear increase with PC period (Fig. 4(b)), showing excellent agreement with the theoretically calculated wavelength via the generalized BR condition. The mean relative error between experimental and theoretical BR wavelengths is approximately 0.09%, demonstrating the high accuracy of the experimental results. The perfect wavelength matching with the Bragg condition provides unambiguous evidence for the lasing from BR mode under TM polarized pump. The polarization-switchable behaviors can directly be observed in the measured angle-resolved PL spectra for different periods of 363 nm (green star), 367 nm (yellow star), and 373 nm (orange star) (Fig. 4(c, d)) and corresponding momentum space images (right inset). These results unambiguously verified the PC period-independent polarization selectivity in the designed DFB lasers.

Moreover, the PC-waveguide-coupled structure with a DC of 0.53 was fabricated as a DFB laser. The typical q-BIC lasing modes are observed under TE polarized wave pump, as well as under TM polarized wave pump (Fig. S7, Supplemental Material). The characteristics of polarization-insistent q-BIC lasing emission is consistent with those in other q-BIC lasers[20,21]. These results highlight the decisive role of the DC in manipulating coupling strength of the two Bloch modes for TE and TM waves, respectively. As predicted, the PC-waveguide-coupled cavity with a small DC enables q-BIC and BR polarized lasing, while the cavity with a large DC supports only q-BIC polarized lasing by tuning the pump beam polarization. Thus, the dynamic emission behavior of the DFB lasers can be controlled via manipulating the DC and pump polarization.

**Summary and discussions**

In summary, a polarization-switchable polymer DFB laser was successfully achieved through manipulating the DC of the PC-waveguide-coupled structure. The switching principle is uncovered through developing the generalized BR condition and tailoring the mode coupling strength in a PC-waveguide-coupled cavity. This principle enables design dundamental for the tunable lasers through tailoring the coupling position and strength. The fabricated polarization-switchable DFB lasers exhibit single q-BIC lasing, q-BIC and BR lasing coexistence, and single BR lasing transitions with varying pump polarization. This polarization-switchable characteristic clarifies the relationship of q-BIC and BR lasing mode through mode coupling and radiation characteristics, also supplies a flexible way to manipulate the lasing polarization. Moreover, the proposed switching principle shows unprecedented robustness in varying waveguide thickness and PC period of distributed feedback (DFB) lasers. Importantly, the achieving method of the polarization-switchable laser is flexible, fabrication easy, low-cost and suit for large scale and q-BIC laser with low index contrast polymer. Although the fabricated DFB microcavities exhibit intrinsic losses, their large-scale architecture enables the simultaneous excitation of up to $10^4$ cells within the pump area, facilitating the lasing operation for both q-BIC and BR modes. As a result, the achieved polymer DFB lasers possess the characteristics of low-threshold, high polarization degree, dynamic switchability via pump polarization. The results not only offer a direct view to uncover fundamental operation mechanisms in microcavity lasers, but also provides a simple yet effective approach for developing tunable lasers in integrated photonics, sensing, and information optics.

**Method**

**Theoretical model for PC-waveguide-coupled structure**

To elucidate the interaction mechanisms between the resonance modes in 1D PC and waveguide coupled structure, we have developed a generalized theory to analyze the dispersion relation without the coupling.

For a PC-waveguide-coupled structure, the refractive index of the polymer is denoted as $n_{\text{WG}}$, the thickness of the waveguide is $T$, and the lattice constant of the PC is $\Lambda$. In the PC-waveguide-coupled structure, the electromagnetic modes are quantized into different band branches named as the $\text{TE}_m^l$ for TE polarized waves and $\text{TM}_m^l$ for TM polarized waves due to the in-plane period boundary of the PC and the localization of the waveguide along $z$ direction. $l$ indicates the diffraction order induced by the PC through the Bloch condition along $x$ direction ($l = \pm1, \pm2, ......$). And the waveguide further discretizes the $l$-th order diffraction branch into several branches with $m$ representing the resonance order of the waveguide structure ($m$=0,1, 2, ……) along $z$ direction.

For the modes of $\text{TE}_m^l$ (or $\text{TM}_m^l$), the incident wave vector $\vec{k}_{\text{in}}$ propagating at an angle $\theta$ relative to the $z$-axis in the waveguide, its $z$-component in the waveguide satisfies the resonance condition:

$$\vec{k}_{\perp \text{in}} = \frac{m\pi + \Delta\Phi}{T} \qquad (1)$$

where, $\Delta\Phi$ denotes the negligible phase difference. The in-plane components of the wave vector $\vec{k}_{\text{in}}$ in the XY plane satisfy the interface momentum matching condition:

$$\vec{k}_{\|in} - l\vec{G}_x = \vec{k}_{\|out} \tag{2}$$

Here, $\vec{k}_{\|in}$ denotes the in-plane wave vector component of the incident light in the waveguide, which satisfies $\vec{k}_{\|in} = \sqrt{\vec{k}_{in}^2 - \vec{k}_{\perp in}^2} = n_{WG} \cdot k_0 \sin\theta$. $\vec{G}_x$ represents the PC wave vector, which is periodic along the $x$ direction, satisfying the relationship $\vec{G}_x = \frac{2\pi}{\Lambda}\hat{x}$. $\vec{k}_{\|out}$ is the in-plane wave vector component of the emitted mode, satisfying $\vec{k}_{\|out} = k_0 \sin\varphi$ with the irradiation angle $\varphi \in \left(-\frac{\pi}{2}, \frac{\pi}{2}\right)$ in air. $k_0 = \frac{2\pi}{\lambda_0}$ is the wave vector in free space.

Accounting for the waveguide thickness $T$, the relationship between the free-space wavelength $\lambda_0$ and the resonance modes $TE_m^l$ (or $TM_m^l$) is governed by:

$$\left(n_{WG} \cdot \frac{2\pi}{\lambda_0}\right)^2 - \left(\frac{m\pi}{T}\right)^2 = \left(l\frac{2\pi}{\Lambda}\right)^2 + \left(\frac{2\pi}{\lambda_0} \cdot \sin\varphi\right)^2 + 2l\frac{2\pi}{\Lambda}\frac{2\pi}{\lambda_0} \cdot \sin\varphi \tag{3}$$

When the waveguide thickness can be neglected, the interface momentum matching condition can be expressed as:

$$\Lambda(n_{WG} \cdot \sin\theta - \sin\varphi) = l \cdot \lambda_0 \tag{4}$$

Given the vertical-cavity surface-emitting characteristic of the DFB microcavity, where the emission angle $\varphi$ is zero, Eq. (3) degenerates to the traditional BR condition with the diffraction order $l = \pm 1$[43]. Thus, Eq. (3) can be looked as the generalized BR condition with the diffraction order $l = \pm 1$.

For the PC-waveguide-coupled cavity, Eq. (3) is simplified to the standard form of a quadratic equation:

$$A\lambda_0^2 + B\lambda_0 + C = 0 \tag{5}$$

where, $A = \frac{l^2}{\Lambda^2} + \frac{m^2}{4T^2}$, $B = \frac{2l\sin\varphi}{\Lambda}$, $C = \sin^2\varphi - n_{WG}^2$.

The solutions to the Eq. (5) can be formulated as:

$$\lambda_0 = \frac{-B + \sqrt{B^2 - 4AC}}{2A} \tag{6}$$

This formula supplies the dispersion relation of the 1D PC-waveguide-coupled structure, indicating the origin of each band branch $TE_m^l$ (or $TM_m^l$) from the $l$-th diffraction order and the $m$-th Bloch modes.

**Full-wave simulation**

In the 1D PC-waveguide-coupled structure, R6G-doped photoresist material ($n$=1.64) was used for both the PC and waveguide part on glass ($n$=1.5) substrate to form a DFB microcavity with a low index contrast. In the constructed model, the PC layer thickness ($D$) is set at 290 nm, where the lattice period ($\Lambda$) and duty cycle (DC = $w/\Lambda$, with $w$ representing the width of medium pillar) serve as key parameters for controlling mode coupling characteristics. The waveguide thickness ($T$) represents another essential parameter that directly influences the dispersion properties of Bloch modes.

For numerical simulations of PC-waveguide-coupled structure, three-dimensional full-wave simulations using the finite-element method is employed. In the 2D modeling framework, periodic boundary conditions are imposed along the x directions. To calculate eigenmodes, perfectly matched layers are applied in the z direction for effective radiation absorption. The mode analysis modules are utilized to solve for dispersion relations, electromagnetic field distributions, and $Q$ factors of the PC-waveguide-coupled structure. The quality factor is determined by $Q=(\text{Re}[\nu])/(2\text{Im}[\nu])$, where Re[$\nu$] and Im[$\nu$] denote the real and imaginary parts of the eigenfrequency of a simulated cavity eigenmode, respectively. The research initially examines the influence of DC variations within the range of 0.21 to 0.68 on momentum-space localization characteristics of coupled modes at a fixed conditions of $\Lambda$=380 nm and $T$=1.64 μm. The optimized DC of 0.34 demonstrates distinct momentum-space localization behaviors for TE and TM waves. Further investigation maintains this optimal DC while varying waveguide thickness (0.90 to 1.92 μm) and PC period (363 to 386 nm) to analyze the stability and robustness of momentum-space localization for different coupled modes.

**Fabrication and Optical Measurements**

The photoresist (AZ® MiR™ 701 Series) films doped with R6G were deposited on the glass substrate through the spin-coating method. The 1D PC with different period was then fabricated on surface of the film by using a dual-beam interference lithography technique. A continuous-wave laser (He-Cd Laser, CW, 325 nm) was employed as the interference light source. The DC of the PC was tuned by controlling the exposure dose and development time.

All optical measurements were performed using a home-made angle-resolved imaging and spectroscopy system designed with special optical lens configurations (Fig.

S3, Supplementary Information). The system features exceptional angular resolution capability, achieving a minimum angular spacing of 0.002 rad and a numerical aperture (NA) of 0.22. A nanosecond laser (MCA-532-1-60-01-PD) was employed as the pump source, with a repetition rate of 500 Hz and a central wavelength of 532 nm. The pump energy density was controlled by a half-wave plate before a linear polarizer which was used to control the polarization state of the pump beams. The polarization state of the emitted light was analyzed using a linear polarizer. The angle-resolved transmission spectra and photoluminescence spectra were obtained by using a spectrometer (Andor-SR-500i-D1-R) with a 2D CCD array at back focal plane. The real-space and momentum-space images were recorded by the camera (RH-CAM4K8MPA).

**Supporting Information**

The Supporting Information is available free of charge via the journal's website.

**Acknowledgments**. National Natural Science Foundation of China (grant no. 92150109 and 61975018).

**Disclosures**. The authors declare no conflicts of interest.

**Data availability**. Data underlying the results presented in this paper are not publicly available at this time but may be obtained from the authors upon reasonable request.

**Figure Captions:**
**Figure 1**

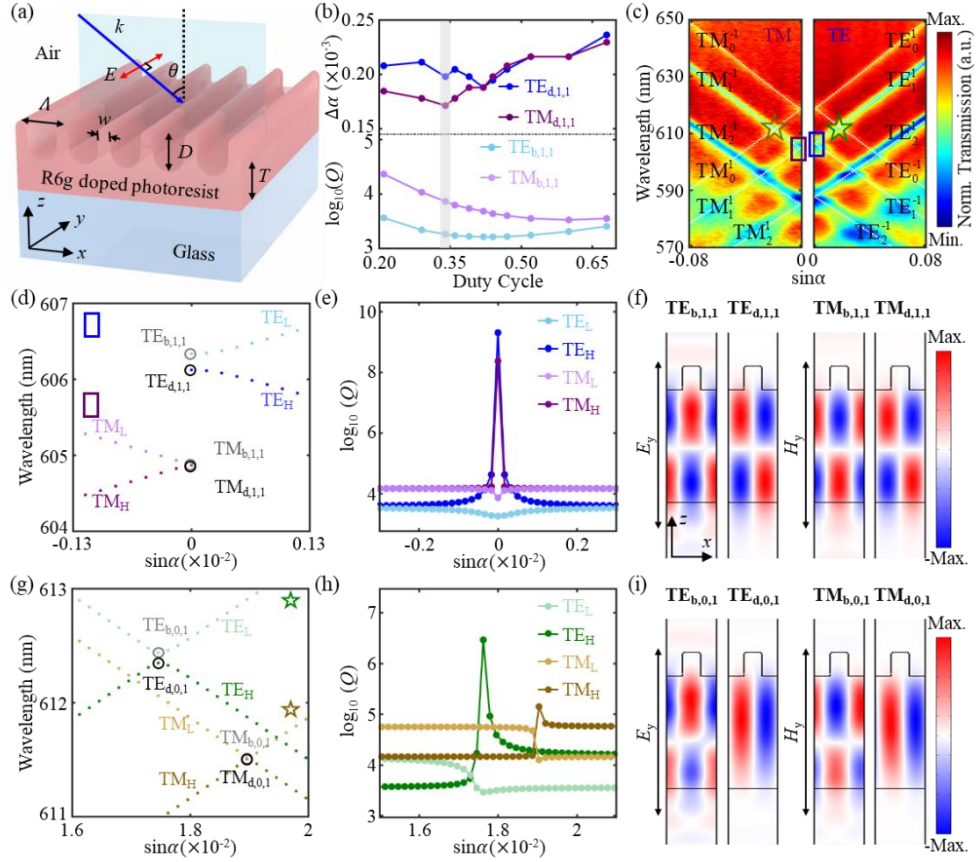

**Fig. 1** (a) Schematic of a 1D PC-waveguide-coupled structure as DFB microcavity. (b) The $Q$-factor angular width for the dark mode (upper panel), and $Q$ value for the bright mode (lower panel) vary with DC of the PC. (c) The experimental angle-resolved transmission spectra and the simulated dispersion curves (white dashed lines) for TM (left panel) and TE (right panel) waves (DC=0.34). (d, e) The local dispersion curves (d), and $Q$-factor curves (e) along the ΓX direction near the Γ points corresponding to the marked blue and purple hollow rectangles in **c**. (f) The distributions of electric field component $E_y$ and magnetic field component $H_y$ of BIC and bright modes at the Γ points, respectively. (g, h) The local dispersion curves (g), and the $Q$-factor curves (h) of the coupled modes from inter-order Bloch mode coupling ($m$ =0 and $m'$=1) near sin$α$ = 0.017 (green pentagram in **c**) for TE waves and sin$α$ = 0.019 (brown pentagram in **c**) for TM waves. (i) The distributions of electric field component $E_y$ and magnetic field component $H_y$ of BIC and bright modes at the off-Γ high-symmetry position, respectively.

**Figure 2**

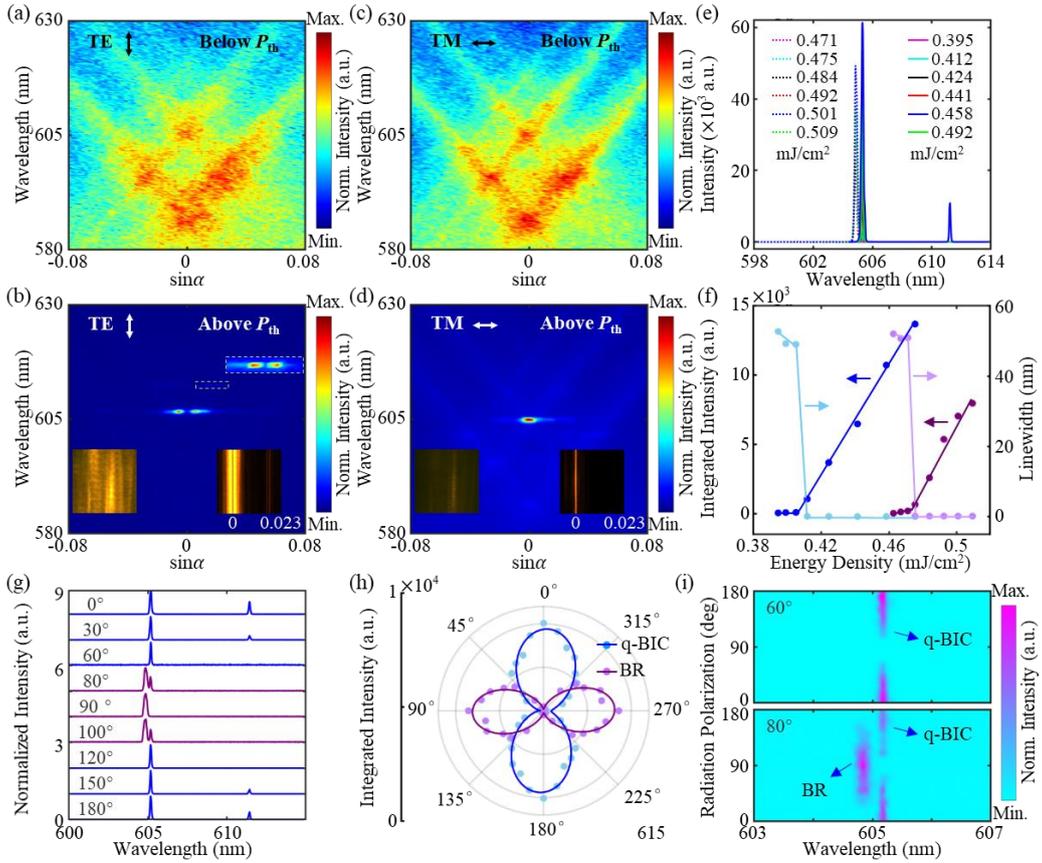

**Fig. 2** (a-d) Normalized angle-resolved PL spectra of the DFB laser under TE (a, b) and TM (c, d) polarization excitation, showing different emission properties below (a, c) and above (b, d) the corresponding threshold. The q-BIC lasing modes from inter-order Bloch mode coupling are observed near sin$\alpha$ = 0.017 under TE polarization pump (white dashed-line box in **b**) as predicted. The insets are the corresponding real-space image (left) and momentum-space image (right) above threshold. (e) The emission spectra vary with the energy density under TE (solid lines) and TM (dashed lines) polarization pump. (f) The threshold behavior of q-BIC lasing under TE polarization pump (dark blue and light blue) and BR lasing under TM polarization pump (dark purple and light pink). (g) The normalized emission spectra vary with pump polarization angle $\theta_p$. (h) Orthogonal polarization characteristics of the q-BIC lasing under TE polarization ($\theta_p$=0°, dark blue) pump and BR lasing under TM polarization ($\theta_p$=90°, dark purple) pump. (i) Radiation polarization characteristics of the emitted lasing at the pump polarization angles of $\theta_p$=60° (top panel) and $\theta_p$=80° (lower panel).

Figure 3

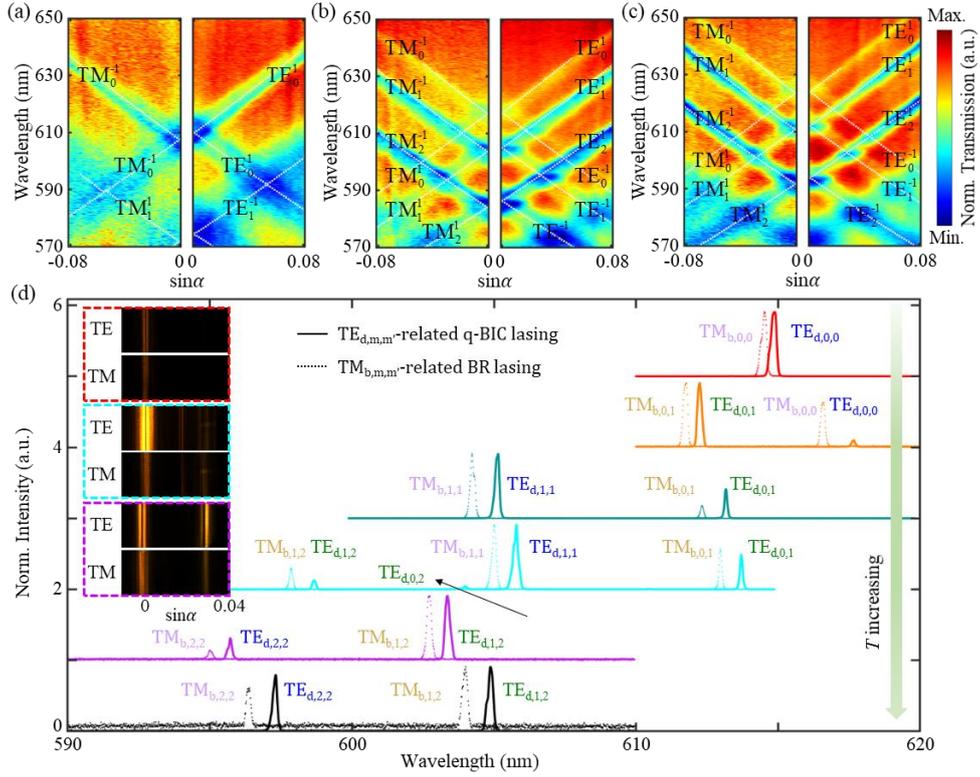

**Fig. 3** (a-c) The measured angle-resolved transmission spectra and the simulated dispersion curves (white dashed lines) for TM waves (left panel) and TE waves (right panel) at a thickness of $T = 0.90$ μm (b), $T = 1.65$ μm (c), and $T = 1.90$ μm (c). (d) Normalized PL spectra of q-BIC lasing (solid lines) and BR lasing (dashed lines) in DFB lasers with different thicknesses. The insets are the emission patterns in momentum space for the DFB lasers with $T = 0.90$ μm (red dashed boxes), $T = 1.65$ μm (cyan dashed boxes), and $T = 1.90$ μm (purple dashed boxes).

**Figure 4**

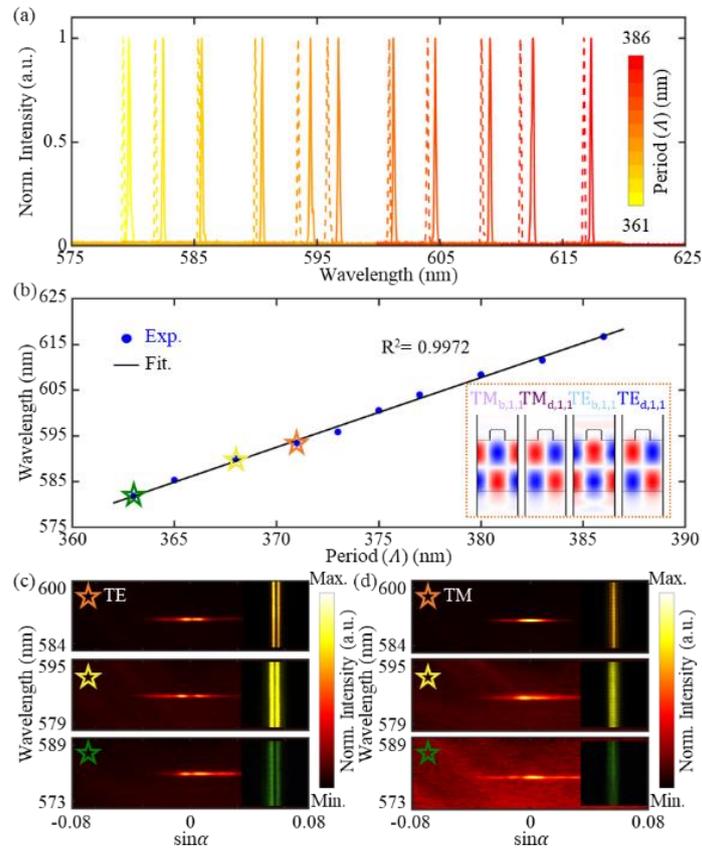

**Fig. 4** (a) The normalized q-BIC lasing (solid lines) and BR lasing (dashed lines) from the obtained DFB lasers with different PC periods $\Lambda$. (b) The experimental radiation wavelength of BR lasing varies with $\Lambda$, as predicted (black solid line). The inset shows the field distribution in the DFB microcavity with $\Lambda$ = 373 nm shown in the orange dashed box. (c, d) Angular-resolved PL spectra for q-BIC lasing (c) and BR lasing (d) at the periods of $\Lambda$ = 363 nm, $\Lambda$ = 367 nm, and $\Lambda$ = 373 nm. The insets are the corresponding momentum-space radiation patterns.